\documentclass{ecai}
\usepackage{times}
\usepackage{graphicx}
\usepackage{latexsym}
\usepackage{hyperref}
\usepackage{amsmath}

\begin{document}

\title{Inter-slice image augmentation based on frame interpolation for boosting medical image segmentation accuracy}

\author{Zhaotao Wu\institute{School of Computer Science and Engineering, South China University of Technology, Guangzhou, China. Jia Wei is the corresponding author, email: csjwei@scut.edu.cn}
\and Jia Wei$^1$
\and Wenguang Yuan$^1$ 
\and Jiabing Wang$^1$
\and Tolga Tasdizen \institute{Scientific Computing and Imaging Institute, University of Utah, Salt Lake City, USA, email: tolga@sci.utah.edu}}

\maketitle
\bibliographystyle{plain}

\begin{abstract}
We introduce the idea of inter-slice image augmentation whereby the numbers of the medical images and the corresponding segmentation labels are increased between two consecutive images in order to boost medical image segmentation accuracy. Unlike conventional data augmentation methods in medical imaging, which only increase the number of training samples directly by adding new virtual samples using simple parameterized transformations such as rotation, flipping, scaling, etc., we aim to augment data based on the relationship between two consecutive images, which increases not only the number but also the information of training samples. For this purpose, we propose a frame-interpolation-based data augmentation method to generate intermediate medical images and the corresponding segmentation labels between two consecutive images. We train and test a supervised U-Net liver segmentation network on SLIVER07 and CHAOS2019, respectively, with the augmented training samples, and obtain segmentation scores exhibiting significant improvement compared to the conventional augmentation methods.

\end{abstract}

\section{INTRODUCTION}
%
Data augmentation has become an indispensable method to boost the performance of state-of-the-art machine learning approaches, especially when the amount of training samples is relatively small. Whereas conventional data augmentation methods in medical imaging typically increase the number of training samples directly by adding new virtual samples using simple parameterized transformations, we introduce the idea of augmenting data based on the relationship between two consecutive images, which increases not only the number but also the information of training samples. Usually, medical volume data have high resolution in the in-plane direction and low resolution in the through-plane direction, because the inter-slice distance is much larger than the x-y spacing in the in-plane direction. With this characteristic, we aim to generate the intermediate slice between two consecutive slices in the through-plane direction. Considering that videos are a continuous image sequence in time, in this paper, we assume that medical volume data are a continuous image sequence in space. With the similarity of videos and medical volume data, we can make use of the idea of frame interpolation to generate intermediate slice between two consecutive slices in the through-plane direction.  

In this paper, we propose a method that generates synthetic inter-slice images based on frame interpolation and an attention mechanism, which can generate an arbitrary number of intermediate medical images from two consecutive images. Our main idea is to warp two consecutive images and the corresponding segmentation labels to the specific space step and then fuse the two warped images and labels to generate the intermediate image and the corresponding label. Inspired by how experienced clinicians segment based on the target object and its surroundings, after generating the intermediate images, we feed the synthetic images and the real images into two discriminators, the global and the local. The global discriminator is used for distinguishing the real and synthetic images. With the global discriminator, the authenticity of the synthetic images can be increased. The local discriminator model is used for distinguishing the part of the image that is useful for segmentation. In the local discriminator model, the attention network is used to automatically focus on the useful features for segmentation. The part that is focused will then be fed into the local discriminator. By using the attention network and the local discriminator, the authenticity of the useful part of the synthetic images can be increased.

The contributions of this paper are threefold. Firstly, to the our best knowledge, no studies have examined medical image data augmentation using the idea of frame interpolation to generate an arbitrary number of intermediate medical images and corresponding segmentation labels from two consecutive images and labels. Secondly, we use two discriminators, the global and the local,  to increase not only the authenticity of the entire image but also the authenticity of the useful part of the image, which can obviously increase the quality of synthetic images. Thirdly, we introduce an adaptive attention network in the local discriminator model that can automatically focus on the part of the image that is useful for classifying and segmenting the target object.

\section{RELATED WORK}
\subsection{Medical image segmentation}
Medical image segmentation tasks have not been satisfactorily solved due to the lack of training data and the large differences in each sample. Researchers have proposed numerous methods to solve the problem of medical image segmentation. These methods can be divided into region-based segmentation methods \cite{kumar2013automatic}, edge-detection-based segmentation methods \cite{mavska2013segmentation}, graph-based segmentation methods \cite{oda2011organ} and deep-learning-based segmentation methods \cite{ciresan2012deep, long2015fully, ronneberger2015u,sandfort2019data}. Deep-learning-based segmentation methods have performed better than the classical segmentation methods and have attained state-of-the-art accuracy, which has attracted the attention of researchers.

In the ISBI 2012 EM Segmentation Challenge, Ciresan et al. \cite{ciresan2012deep} were the first to use CNN \cite{lecun1995convolutional, lecun1998gradient} to segment neuronal membranes in electron microscopy images. In 2015, Long et al. \cite{long2015fully} proposed a fully convolutional network (FCN) and used it in semantic segmentation. Different from CNN, which has convolutional layers and fully connected layers, FCN has only convolutional layers, so that FCN can take input of arbitrary size and produce correspondingly sized output with efficient inference and learning. Ronneberger et al. \cite{ronneberger2015u} proposed U-Net, which is widely used in medical image segmentation. The architecture of U-Net is similar to that of an auto-encoder \cite{masci2011stacked}. Both of them consist of a contracting path and an expansive path. However, different from an auto-encoder, U-Net concatenates the feature map from the contracting path with the corresponding feature map from the expansive path, which can enable the model to learn shallow information such as the position and texture of the image and deep information containing semantics.

\subsection{Data augmentation}
In recent years, the success of deep learning in the image domain has benefited from the powerful capacity of the model and the large amount of data. The huge amount of data improves the generalization ability of the model and avoids model overfitting. Numerous experiments and studies have proven that the most effective way to improve the performance of a model is to collect more high-quality data. However, this method is hard to implement in the field of medical imaging. Problems such as scarce cases, tight medical resources, and expensive labeling have led researchers to turn to how to better use existing data, namely data augmentation. 

Data augmentation is a common way to enrich data in the field of deep learning. Data augmentation methods such as rotation, scaling, translation, gamma correction and elastic transformation \cite{simard2003best} are widely used. These methods are easy to implement and are effective at improving testing performance \cite{oliveira2017augmenting, pereira2016brain, ronneberger2015u, roth2015deeporgan, zhao2019data}. 

Rotation rotates the image in a random angle. Scaling enlarges or reduces the scale of the image, which can improve the ability of the model to segment small targets. Translation involves moving the image in the X or Y direction (or both). Translation is useful because most objects can be located almost anywhere in the image, which can force the model to "look" at any location of the image. Gamma correction performs a nonlinear operation on each pixel of the input image, which can make the pixel of the output image exponentially related to the pixel in the input image. Elastic transformation was first used in the MNIST handwritten digit classification problem \cite{simard2003best}. It first generates two random numbers from the range $[-1,1]$ for each pixel of the image, then generates a Gaussian kernel to convolute with the random number, and finally acts on the original image to create random movement. It is widely used in the data augmentation of medical images.

\subsection{Video interpolation}
Video interpolation is one of the basic video processing techniques used to generate intermediate frames between any two consecutive original frames. Nowadays, with the development of deep learning, deep-learning-based methods are widely used in video interpolation \cite{dosovitskiy2015flownet, Ilg2017FlowNet, ranjan2017optical}. Several deep-learning-based methods have been proposed to predict optical flow with input frames to obtain the interpolated frames. Niklaus et al. \cite{niklaus2018context} proposed a context-aware frame synthesis approach that warps not only the input frames but also their pixel-wise contextual information to interpolate a high-quality intermediate frame. Bao et al. \cite{bao2019depth} developed a depth-aware flow projection layer to synthesize intermediate flows that samples closer objects rather than farther ones. Their method exploits the optical flow, local interpolation kernels, depth maps, and contextual features to synthesize high-quality video frames. Jiang et al. \cite{jiang2018super} proposed an end-to-end convolutional neural network for variable-length multi-frame video interpolation, where the motion interpretation and occlusion reasoning are jointly modeled. They linearly combine the bi-directional optical flow between the input images at each time step to approximate the intermediate bi-directional optical flow, which can help the method generate arbitrary higher frame rate videos. 

Different from the methods described above, several recent methods directly interpolate video frames without estimating optical flow. Simon et al. \cite{niklaus2017video} employed a deep fully convolutional neural network to estimate a spatially adaptive convolution kernel for each pixel. This deep neural network can be directly trained end to end using the input images without any ground truth data such as optical flow, which are difficult to obtain. Inspired by the success of using separable filters to approximate full 2D filters for other computer vision tasks, Simon et al. \cite{niklaus2017videof} also developed a new method that takes two input frames and estimates pairs of 1D kernels for all pixels simultaneously without any optical flow. For frame synthesis, two 2D convolution kernels are required to generate an output pixel. The method approximates each 2D kernels with a pair of 1D kernels, one horizontal and one vertical. Therefore, a $n \times n$ convolution kernel can be encoded using only $2n$ variables, which can significantly decrease the number of parameters. 

\begin{figure*}
\centerline{\includegraphics[scale=0.25]{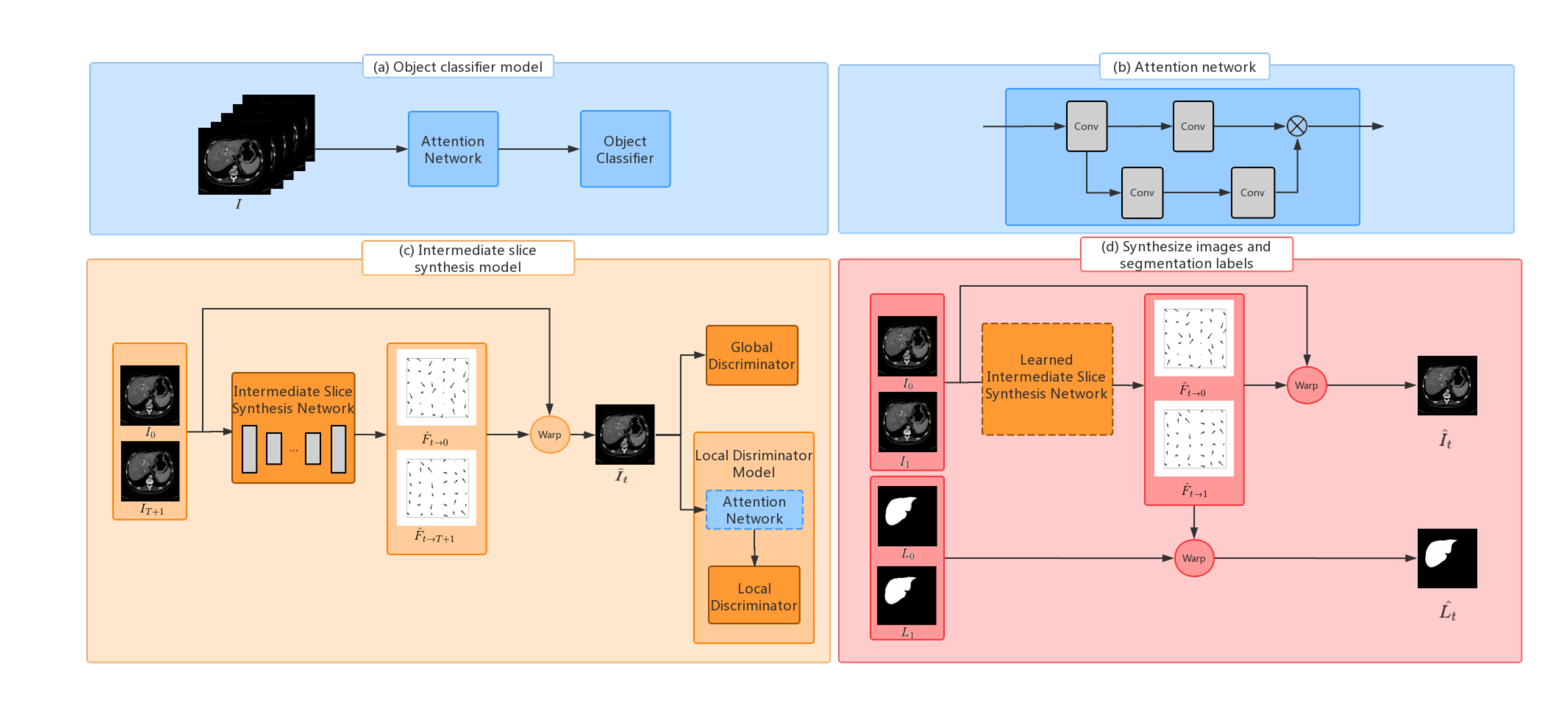}}
\caption{An overview of the proposed method: (a) shows the architecture of the object classifier model, (b) shows the details of the attention network, (c) shows the architecture of the intermediate slice synthesis model, and (d) shows the process of synthesizing inter slice and corresponding segmentation label. The dashed line in the figure indicates that the parameters is frozen.} 
\label{fig2}
\end{figure*}
\subsection{Attention mechanism}

The attention mechanism is widely used in the field of imaging. It allows models to learn deeper correlations between objects \cite{mnih2014recurrent} and helps model discover interesting new patterns of data \cite{hoogi2019self, jo2018quantitative, olshausen1993neurobiological}. Ma et al. \cite{ma2018gan} proposed a framework for instance-level image translation by deep attention GAN. They decompose the task to instance-level image translation to make the controllability more enhanced, and integrate the attention mechanism into a GAN, which makes the network automatically and adaptively learn task-driven identity representations without human involvement. Liu et al. \cite{liu2019end} proposed a multi-task learning architecture that allows learning of task-specific feature-level attention. They use a single shared network to calculate the global feature, together with a soft-attention module for each task. The soft-attention modules make the network learn the task-specific features from the global features. The features can be shared among different tasks at the same time. Chen et al. \cite{chen2019multi} proposed a semi-supervised image segmentation method that simultaneously optimizes a supervised segmentation and an unsupervised reconstruction objective. The reconstruction objective uses an attention mechanism that separates the reconstruction of image areas corresponding to different classes. By using the attention mechanism, their method is encouraged to learn discriminative features of segmentation from unlabeled images. 

\section{PROPOSED APPROACH}
We propose a method to boost medical image segmentation accuracy by synthesizing realistic training examples based on frame interpolation and an attention mechanism. The proposed method is summarized in Figure \ref{fig2}. 

First of all, the object classifier, which is used for distinguishing whether the input image has the target object, is trained. It is made by an attention network and a classifier, shown in Figure \ref{fig2}(a). The details of the attention network are shown in Figure \ref{fig2}(b). The attention network has two branches: one branch extracts the image's feature, and the other branch generates the corresponding activation map. The input images $I$ are first fed into the attention network, and then the attention network outputs the features that are elements-wise multiplied by the extracted features and the activation maps. The output features are then fed into the object classifier. 

After training the object classifier, the intermediate slice synthesis network is trained to generate intermediate slices at any space step between two input images. We divide the training data into multiple sets in sequence, each with $T+2$ images, where $T$ denotes the number of the intermediate slices. Given two input images $I_{0}$ and $I_{T+1}$, what we want to do is to predict the intermediate slices, $\{\hat{I}_{t}\}_{t=1}^{T}$, which should be as close as possible to the ground truth intermediate slices $\{I_{t}\}_{t=1}^{T}$. The details of the intermediate slice synthesis network are shown in Figure \ref{fig2}(c). The intermediate slice synthesis network generates the spatial transformations $\hat{F}_{t\rightarrow0}$ and $\hat{F}_{t\rightarrow T+1}$. $I_{0}$ and $I_{T+1}$ are warped by $\hat{F}_{t\rightarrow0}$ and $\hat{F}_{t\rightarrow T+1}$ to generate the synthetic intermediate slice $\hat{I}_{t}$. Then $\hat{I}_{t}$ and $I_{t}$ are fed into the local discriminator model and the global discriminator. Note that the local discriminator model contains an attention network and a discriminator. The attention network is pre-trained in the object classifier and is frozen when the local discriminator model is trained. 

Finally, the learned intermediate slice synthesis network is used to generate synthetic images and corresponding segmentation labels between two consecutive input images. Let $I_{0}$ and $I_{1}$ denote the input consecutive images and let $L_{0}$ and $L_{1}$ denote the corresponding segmentation labels, while $\hat{I}_{t}$ and $\hat{L}_{t}$ denote the synthetic images and segmentation labels at space step $t\in(0,1)$. $I_{0}$, $I_{1}$ and $L_{0}$, $L_{1}$ are warped by the same spatial transformations $\hat{F}_{t\rightarrow0}$ and $\hat{F}_{t\rightarrow 1}$to generate $\hat{I}_{t}$ and $\hat{L}_{t}$.The process of synthesizing inter slices and corresponding segmentation labels is shown in Figure \ref{fig2}(d).  

\textbf{Object classifier model}: 
In order to train an attention network that can automatically focus on the target object and other useful information for classification, we first train an object classifier model that contains an attention network to distinguish whether the input image has the target object. By doing so, the attention network can automatically focus on the part of the image that is useful for classification. The loss function that we train the object classifier is shown as follows:

\begin{equation}
l = - \sum_{i=1}^{n} Y_{i} \log \hat{Y}_{i}
\label{eq6_1}
\end{equation}
where $n$ denotes the size of the batch, $Y_{i} \in \{0,1\}$ denotes whether the input image has the target object, $Y_{i} = 1$ if the input image has the target object and $Y_{i} = 0$ otherwise. $\hat{Y}_{i} \in \{0,1\}$ denotes the output given by the classifier whether the input image has the target object, it has the same definition as $Y_{i}$.

\textbf{Intermediate slice synthesis model}: 
The intermediate slice synthesis model is similar to the method proposed by Jiang et al. \cite{jiang2018super}. We compute the bi-directional spatial transformations between the input images using a U-Net architecture. The input images are warped by the spatial transformations to generate the intermediate slice according to the following equations: 

\begin{equation}
\hat{I_{t}} = (1-\frac{t}{T+1}) g(I_{0}, \hat{F}_{t\rightarrow0}) + \frac{t}{T+1} g(I_{T+1}, \hat{F}_{t\rightarrow T+1})
\label{eq1}
\end{equation}

where $g(\cdot, \cdot)$ is a backward warping function, which is implemented using bilinear interpolation \cite{liu2017video, zhou2016view}.

The loss function of the intermediate slice synthesis network is shown as follows:

\begin{equation}
\begin{aligned}
l =& \lambda_{rec} l_{rec} + \lambda_{per} l_{per} + \lambda_{warp} l_{warp} \\
&+ \lambda_{smooth} l_{smooth} + \lambda_{adv} l_{adv}
\end{aligned}
\label{eq7}
\end{equation}

Equation \ref{eq7} is a linear combination of five terms, where $\lambda$ is the weight of each term to control the contribution of each term. The reconstruction loss $l_{rec}$, the perceptual loss $l_{per}$, the warping loss $l_{warp}$ and the smoothness loss $l_{smooth}$ are defined in the method proposed by Jiang et al. \cite{jiang2018super}.

The fifth term of Equation \ref{eq7} is $l_{adv}$. It is the adversarial loss, which encourages the generator to synthesize the image to confuse the two discriminators. It can improve the authenticity of the synthetic images. It is shown as follows:

\begin{equation}
l_{adv} = -\frac{1}{T} \sum_{t=1}^{T}\log LD(\hat{I}_{t}) - \frac{1}{T} \sum_{t=1}^{T}\log GD(\hat{I}_{t})
\label{eq12}
\end{equation}
where $LD$ denotes the local discriminator model and $GD$ denotes the global discriminator.

\textbf{Discriminator}: 
In this method, two discriminators, the global and the local, are used to battle with the generator, the intermediate slice synthesis network. The global discriminator is used for distinguishing the synthetic from the real images. The local discriminator is used for distinguishing the useful part of the synthetic images, which are useful for segmentation, from the corresponding part of the real images. In this method, we assume that the parts in the image that contribute to the classification of the target object are also useful for the segmentation of the target object. Therefore, to determine the useful parts of the images, the attention network, which comes from the object classifier model, is used.  

The global discriminator is fed by the whole image, and the loss function used to optimize is shown as follows:

\begin{equation}
l_{global} = - \frac{1}{T} \sum_{t=1}^{T}\log (1 - GD(\hat{I}_{t})) - \frac{1}{T} \sum_{t=1}^{T}\log GD(I_{t})
\label{eq13}
\end{equation}

The loss function for the local discriminator is shown as follows:

\begin{equation}
l_{local} = - \frac{1}{T} \sum_{t=1}^{T}\log (1 - LD(\hat{I}_{t})) - \frac{1}{T} \sum_{t=1}^{T}\log LD(I_{t})
\label{eq14}
\end{equation}

\textbf{Synthesizing images and segmentation labels}: 
The learned intermediate slice synthesis model is used to generate the intermediate slices and segmentation labels. Similar to the process of training the intermediate slice synthesis network, the two input images are warped by the intermediate bi-directional spatial transformations and are linearly fused to form each intermediate slice. In order to ensure the newly synthesized image is correctly labeled, we use the same spatial transformation to generate the image and its label. The input images and their segmentation labels are warped by the spatial transformations to generate the intermediate slice and its segmentation label according to the following equations: 

\begin{equation}
\hat{I_{t}} = (1-t) g(I_{0}, \hat{F}_{t\rightarrow0}) + t g(I_{1}, \hat{F}_{t\rightarrow 1})
\label{eq3}
\end{equation}

\begin{equation}
\hat{L_{t}} = (1-t) g(L_{0}, \hat{F}_{t\rightarrow0}) + t g(L_{1}, \hat{F}_{t\rightarrow 1})
\label{eq2}
\end{equation}

\section{EXPERIMENT}
\subsection{Datasets}
Two datasets are used in our experiments, one of which is SLIVER07 \cite{van20073d}, which is publicly available through the MICCAI 2007 Segmentation of the Liver challenge \cite{goldman2008principles}. All the images in SLIVER07 are taken from the axial direction, and there is no overlap between consecutive slices. The images of SLIVER07 are generated by a variety of different scanning devices with x-y spacing ranging from 0.55mm to 0.80mm and inter-slice distance ranging from 1mm to 3mm. SLIVER07 contains CT images of 30 patients, and each patient contains from 64 to 394 slices. In order to show the effect of our method when the dataset is small, we use 5 patients, 7 patients and 9 patients, respectively, to train the segmentation network, and we use 5 patients as the test set. The other dataset is CHAOS2019 \cite{ali_emre_kavur_2019_3431873}, which is the challenge of Combined (CT-MR) Healthy Abdominal Organ Segmentation. The images of CHAOS2019 are generated by three different scanning devices with x-y spacing ranging from 0.7mm to 0.88mm and inter-slice distance ranging from 3mm to 3.2mm. CHAOS2019 contains CT images of 40 patients, each patient contains from 77 to 105 slices. In order to show the effect of our method when the dataset is small, we use 5 patients, 7 patients and 9 patients, respectively, to train the segmentation network, and use 5 patients as the test set. The details of the two datasets are summarized in Table \ref{table1}.

\begin{table}
\begin{center}
\caption{Details of the datasets}\label{table1}
\scalebox{1.2}{
\begin{tabular}{lcc}
\hline
\rule{0pt}{12pt}
\rule{0pt}{12pt}
& SLIVER07 & CHAOS2019
\\
\hline
\\[-6pt]
slice per patient & 64-394 & 77-105\\
resolution & $512\times512$ & $512\times512$\\
x-y spacing(mm) & 0.55-0.80 & 0.70-0.80\\
inter-slice distance(mm) & 1.0-3.0 & 3.0-3.2
\\
\hline
\end{tabular}}
\end{center}
\end{table}

\subsection{Implementation}
\label{imple}
In our method, the object classifier model contains an attention network and an object classifier. For the two branches of the attention network, one branch uses one convolution layer to extract the image's feature, and the other branch uses two convolution layers to generate the corresponding attention mask. The object classifier contains only two fully connected layers. The object classifier model is trained in 100 epochs using the learning rate of 0.0005. The batch size $n$ is 6. 

When training the intermediate slice synthesis network, we divide the training data into multiple sets in sequence, each with $T+2$ images, in this paper $T$ is set to 3. The weights of Equation \ref{eq7} are set as $\lambda_{rec} = 2$, $\lambda_{per} = 0.005$, $\lambda_{warp} = 1$, $\lambda_{smooth} = 1$, $\lambda_{adv} = 1$. The attention network in the local discriminator model is pre-trained in the object classifier model and is frozen when the local discriminator model is trained. All the discriminators in our method contain three convolution layers, a fully connected layer and a sigmoid function. The intermediate slice synthesis network is trained in 200 epochs using the base learning rate of 0.0005, which is then reduced by a factor of 10 after the 100th and 150th epochs, and the batch size is set to 6. We adopt an Adam optimizer \cite{kingma2014adam} to optimize all the networks. 

When augmenting and segmenting, the intermediate slice synthesis network first generates synthetic images and corresponding segmentation labels according to Equation \ref{eq3} and Equation \ref{eq2}, and then a U-Net is trained on both the synthetic and real images. Finally, the trained U-Net is used to segment the testing samples. The U-Net we used consists of an encoder and a decoder. The encoder and the decoder have skipped connection at the same spatial. The encoder has four convolution layers. All the convolution layers are followed with an average pooling layer. The decoder has five convolution layers. All the convolution layers except the last layer are followed with a bilinear upsampling layer. The U-Net is trained on 20 epochs using the base learning rate of 0.0001, and the batch size of training is 6. An Adam optimizer \cite{kingma2014adam} is used to optimize the network. The libraries we have used are pytorch0.4.0.

\subsection{Baselines}
In the testing stage, U-Net, whose architecture has been described in Section \ref{imple}, is used to segment the testing data. The baselines used in our experiments are shown as follows:

\textbf{Without data augmentation ($normal$)}: The U-Net is trained with the original data without data augmentation. This method serves as a lower bound.

\textbf{Rotation ($rotation\textendash aug$)}: In our experiments, we rotate each image $90^{\circ}$, $180^{\circ}$ and $270^{\circ}$, respectively, to form the new image.

\textbf{Scaling ($scaling\textendash aug$)}: In our experiments, we randomly change the scale of the image in the range of $[0.8,1.25]$, and each image is scaled three times.

\textbf{Gamma correction ($gamma\textendash aug$)}: In our experiments, the gamma value is set in the range of $[0.7,1.5]$, and each images uses the gamma correction three times. 

\textbf{Elastic transformations ($rand\textendash aug$)}: Elastic transformation is the most common method used in medical image data augmentation. This method enriches the dataset by applying spatial deformations to each image, called elastic transformations \cite{simard2003best}. It first generates two random numbers from the range $[-1,1]$ for each pixel of the image, then generates a Gaussian kernel to convolute with the random number, and finally acts on the original image to make random movement. In our experiments, each image uses the elastic transformation three times.

\subsection{Variants of our method}
\textbf{Data augmentation using intermediate slice synthesis ($ours\textendash normal$)}: Only the intermediate slice synthesis network is used, without using the two discriminators. In the experiments, three images are interpolated between two consecutive images. 

\textbf{Data augmentation using intermediate slice synthesis with two discriminators ($ours\textendash Dis$)}: To highlight the effect of the two discriminators in our method, the intermediate slice synthesis network is used to battle with the global discriminator and the local discriminator model. The local discriminator model in this method does not use the attention network and has the same architecture as the global discriminator. In order to replace the attention mechanism in the local discriminator model, the image is multiplied by its segmentation label to generate the attention image. The images without liver will be input only to the global discriminator. In our experiments, three images are interpolated between two consecutive images.

\textbf{Data augmentation using intermediate slice synthesis with two discriminators and an attention mechanism ($ours\textendash Dis\textendash Att$)}: To highlight the efficacy of the attention mechanism in our method, we add the attention network in the local discriminator model. Similar to ${ours\textendash Dis}$, the intermediate slice synthesis network is used to battle with the local discriminator model and the global discriminator. In the experiments, three images are interpolated between two consecutive images.

\subsection{Evaluation metrics}
The Dice score \cite{dice1945measures} is used to evaluate the accuracy of the U-Net with different data augmentation methods. The formulation of the Dice score is shown as follows:

\begin{equation}
Dice(M, \hat{M}) = 2\times (\frac{|M\cap \hat{M}|}{|M| + |\hat{M}|})
\label{eq15}
\end{equation}
where $M$ denotes the ground truth of the real image and $\hat{M}$ denotes the predicted segmentation label. The Dice score quantifies the overlap between two segmentation labels. If the Dice score is 0, the two labels have no overlap. With the Dice score increasing, the two labels have more overlap. When the Dice score is 1, the two labels have completely overlap. The Dice score is used in the field of medical image segmentation and is one of the most commonly used methods of evaluating segmentation accuracy.

\subsection{Results}
\subsubsection{Synthesized images}
\begin{figure}
\centerline{\includegraphics[scale=0.43]{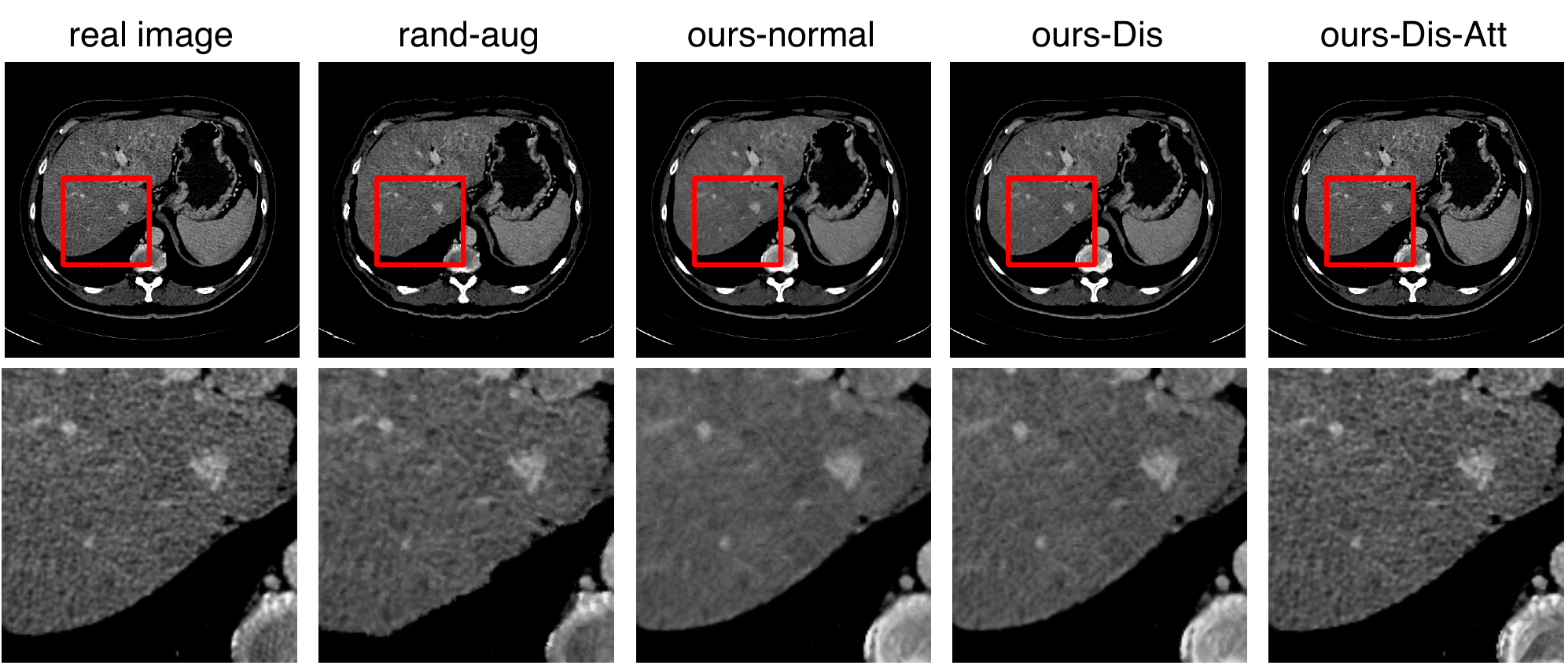}}
\caption{Synthetic images of different methods in SLIVER07.} 
\label{fig2_5}
\end{figure}

\begin{figure}
\centerline{\includegraphics[scale=0.43]{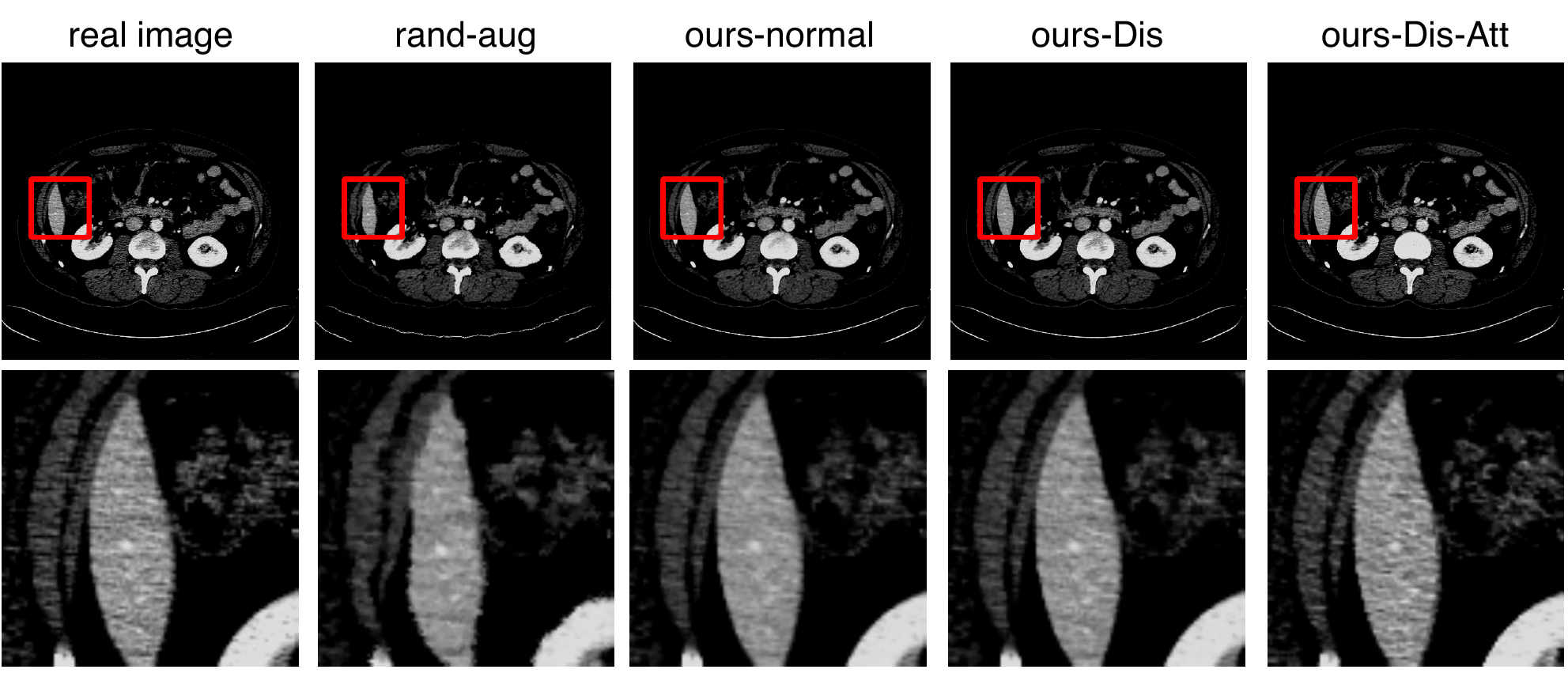}}
\caption{Synthetic images of different methods in CHAOS2019.} 
\label{fig3}
\end{figure}

Synthetic images of some methods in SLIVER07 and CHAOS2019 are shown in Figure \ref{fig2_5} and Figure \ref{fig3}. Considering that $rotation\textendash aug$, $scaling\textendash aug$ and $gamma\textendash aug$ do not change the texture of the images, we show only the real image and the synthetic images of $rand\textendash aug$, $ours\textendash normal$, $ours\textendash Dis$ and $ours\textendash Dis\textendash Att$. From Figure \ref{fig2_5} and Figure \ref{fig3}, we can see that the edge of the synthetic image using elastic transformations (column 2) is not smooth. One possible reason is that elastic transformation makes random movement on the original image to generate the new image, which will make some sharp angle in the edge of the synthetic image. By contrast, our method (column 3 to 5) can generate the image with a more smooth edge, which means the synthetic image is more realistic than the image generated by elastic transformations. In Figure \ref{fig2_5} and Figure \ref{fig3}, $ours\textendash Dis\textendash Att$ (column 5) has the more clear texture in the object than $ours\textendash normal$ (column 3) and $ours\textendash Dis$ (column 4), which means the synthetic image is more realistic than the image generated by $ours\textendash normal$ (column 3) and $ours\textendash Dis$ (column 4) in the target object. This result demonstrates that the two discriminators with the attention mechanism can improve the authenticity of the synthetic image.

For training the network and generating new slices, we use a RTX 2080 Ti GPU to run the model. When training the network with SLIVER07, it takes 45 seconds for each epoch and takes 2 hours and 30 minutes for training the synthetic network. When training the network with CHOAS2019, it takes 68 seconds for each epoch and takes about 3 hours and 42 minutes for training the synthetic network. For generating new slices, if the network interpolates three frames between two images, it takes 36 seconds for generating 300 pictures.

\subsubsection{Segmentation performance comparison with different methods}
\label{section4_3}
Table \ref{table2} and Table \ref{table3} show the segmentation accuracy attained by each method. In most cases, $ours\textendash Dis$ performs better than $ours\textendash normal$, which means the two discriminators can improve the authenticity of the synthetic images because the global discriminator can increase the authenticity of the synthetic images, and the local discriminator based on the label mask can increase the authenticity of the target object in the images. $ours\textendash Dis\textendash Att$ attains the best performance of all the methods, because it uses not only the two discriminators to increase the authenticity of the images but also the attention mechanism to make the local discriminator focus on the target object and its surroundings, which is useful for segmentation.  

\begin{table}
\begin{center}
\caption{Segmentation performance of different methods in terms of the Dice score, evaluated on SLIVER07. We report the mean Dice score (and standard deviation in parentheses) of experiment, which is repeated five times.}\label{table2}
\scalebox{1}{
\begin{tabular}{lccc}
\hline
\rule{0pt}{12pt}
&\multicolumn{3}{c}{Number of patients}\\
\cline{2-4}

\rule{0pt}{12pt}
\rule{0pt}{12pt}
& 5 & 7 & 9 
\\
\hline
\\[-6pt]
$normal$ & $0.555(0.241)$ & $0.711(0.233)$ & $0.738(0.247)$\\
$rotation\textendash aug$ & $0.786(0.094)$ & $0.7902(0.020)$ & $0.818(0.037)$\\
$scaling\textendash aug$ & $0.724(0.198)$ & $0.746(0.128)$ & $0.791(0.205)$\\
$gamma\textendash aug$ & $0.570(0.146)$ & $0.712(0.174)$ & $0.751(0.054)$\\
$rand\textendash aug$ & $0.695(0.037)$ & $0.751(0.0.035)$ & $0.796(0.030)$\\
$ours\textendash normal$ & $0.759(0.019)$ & $0.843(0.019)$ & $0.892(0.016)$\\
$ours\textendash Dis$ & $0.782(0.028)$ & $0.844(0.021)$ & $0.904(0.034)$\\
$ours\textendash Dis\textendash Att$ & $\textbf{0.817(0.018)}$ & $\textbf{0.861(0.030)}$ & $\textbf{0.912(0.012)}$\\
\\
\hline
\end{tabular}}
\end{center}
\end{table}

\begin{table}
\begin{center}
\caption{Segmentation performance of different methods in terms of the Dice score, evaluated on CHAOS2019. We report the mean Dice score (and standard deviation in parentheses) of experiment, which is repeated five times.}\label{table3}
\scalebox{1}{
\begin{tabular}{lccc}
\hline
\rule{0pt}{12pt}
&\multicolumn{3}{c}{Number of patients}\\
\cline{2-4}

\rule{0pt}{12pt}
\rule{0pt}{12pt}
& 5 & 7 & 9 
\\
\hline
\\[-6pt]
$normal$ & $0.635(0.191)$ & $0.663(0.150)$ & $0.706(0.091)$\\
$rotation\textendash aug$ & $0.666(0.064)$ & $0.768(0.086)$ & $0.779(0.142)$\\
$scaling\textendash aug$ & $0.719(0.014)$ & $0.779(0.046)$ & $0.781(0.017)$\\
$gamma\textendash aug$ & $0.656(0.199)$ & $0.755(0.134)$ & $0.764(0.113)$\\
$rand\textendash aug$ & $0.711(0.158)$ & $0.729(0.142)$ & $0.743(0.146)$\\
$ours\textendash normal$ & $0.695(0.086)$ & $0.7951(0.065)$ & $0.807(0.020)$\\
$ours\textendash Dis$ & $0.713(0.124)$ & $0.783(0.102)$ & $0.817(0.021)$\\
$ours\textendash Dis\textendash Att$ & $\textbf{0.733(0.051)}$ & $\textbf{0.814(0.085)}$ & $\textbf{0.843(0.014)}$
\\
\hline
\end{tabular}}
\end{center}
\end{table}

\begin{figure}
\centerline{\includegraphics[scale=0.7]{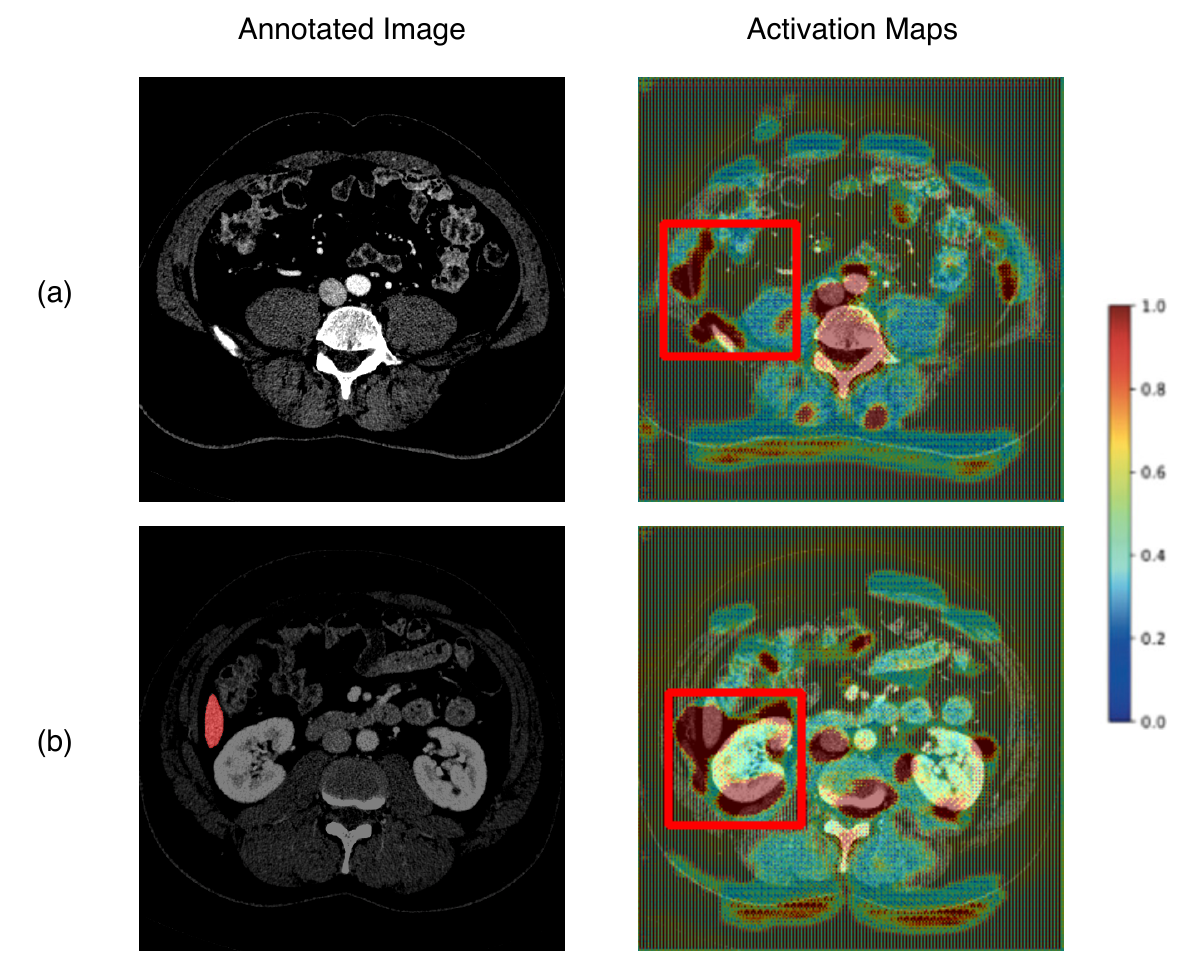}}
\caption{Visualization of activation maps of the image with and without liver: (a) shows the activation map of the image without liver. (b) shows the activation maps of the image with liver (best viewed in color).} 
\label{fig1}
\end{figure}

To show the effect of the attention network in our method, we visualize the activation maps extracted by the attention network. The activation maps of the images are shown in Figure \ref{fig1}. 

From Figure \ref{fig1}(a), although the image does not have liver, the attention network also focuses on the location that is associated with the liver (the red box in Figure \ref{fig1}(a)). From Figure \ref{fig1}(b), it is easy to see that the attention network not only focuses on the liver that we need to segment, but also focuses on some part of the image around the liver (the red box in Figure \ref{fig1}(b)). According to the description above, it is easy to reach the conclusion that the surroundings of the target object are also useful for the classifier to distinguish the target object. Our local discriminator model can find out these useful surroundings.

\subsubsection{Segmentation performance comparison with different numbers of intermediate slices}

In order to show the influence of different numbers of the intermediate slices on our proposed method, we conduct five sets of experiment using the intermediate slice synthesis network to generate different numbers of intermediate slices between two input images. The results are shown in Table \ref{table4} and Table \ref{table5}.

Table \ref{table4} and Table \ref{table5} show that in most cases, when the proposed method interpolates three slices between the two consecutive images, we obtain the best result (row 3). Although the number of enriched images of the method that interpolates three slices is less than the number of enriched images of the method that interpolates four slices and five slices, the performance of segmentation is still the best. One reasonable explanation is that the synthetic images are not real, there will be some deviation between the synthetic images and the real images. So when the number of the synthetic image increases, the deviation between the synthetic images and the real images will increase too. If we interpolate too many slices between the two images, the deviation may increase and finally lead to poor experimental results. 

\begin{table}
\begin{center}
\caption{Segmentation performance of interpolating different numbers of intermediate slices in terms of the Dice score, evaluated on SLIVER07. We report the mean Dice score (and standard deviation in parentheses) of experiment which is repeated five times.}\label{table4}
\scalebox{1.2}{
\begin{tabular}{lccc}
\hline
\rule{0pt}{12pt}
&\multicolumn{3}{c}{Number of patients}\\
\cline{2-4}

\rule{0pt}{12pt}
\rule{0pt}{12pt}
& 5 & 7 & 9 
\\
\hline
\\[-6pt]
1 slice & $0.832(0.039)$ & $0.855(0.031)$ & $0.863(0.046)$\\
2 slices & $0.796(0.039)$ & $0.850(0.022)$ & $0.882(0.012)$\\
3 slices & $0.817(0.018)$ & $\textbf{0.861(0.030)}$ & $\textbf{0.912(0.012)}$\\
4 slices & $\textbf{0.840(0.022)}$ & $0.850(0.034)$ & $0.866(0.021)$\\
5 slices & $0.816(0.107)$ & $0.837(0.017)$ & $0.863(0.019)$
\\
\hline
\end{tabular}}
\end{center}
\end{table}

\begin{table}
\begin{center}
\caption{Segmentation performance of interpolating different numbers of intermediate slices in terms of the Dice score, evaluated on CHAOS2019. We report the mean Dice score (and standard deviation in parentheses) of experiment which is repeated five times.}\label{table5}
\scalebox{1.2}{
\begin{tabular}{lccc}
\hline
\rule{0pt}{12pt}
&\multicolumn{3}{c}{Number of patients}\\
\cline{2-4}

\rule{0pt}{12pt}
\rule{0pt}{12pt}
& 5 & 7 & 9 
\\
\hline
\\[-6pt]
1 slice & $0.690(0.113)$ & $0.785(0.153)$ & $0.802(0.052)$\\
2 slices & $\textbf{0.743(0.098)}$ & $0.798(0.105)$ & $0.826(0.026)$\\
3 slices & $0.733(0.051)$ & $\textbf{0.814(0.085)}$ & $\textbf{0.843(0.014)}$\\
4 slices & $0.733(0.060)$ & $0.811(0.093)$ & $0.810(0.049)$\\
5 slices & $0.734(0.121)$ & $0.799(0.067)$ & $0.831(0.080)$
\\
\hline
\end{tabular}}
\end{center}
\end{table}



\section{Conclusion}
In this paper, we have proposed a data augmentation method that is based on frame interpolation and an attention mechanism for boosting medical image segmentation accuracy. Our method can generate as many intermediate medical images as needed between two input consecutive images. The experiments comparing different methods on SLIVER07 and CHAOS2019 demonstrate the effect of our method. The ablation experiments demonstrate the effectiveness of the two discriminators and the attention mechanism, and show that in most cases, when the network interpolates three slices between two images, the segmentation network can receive the best result. In this paper, we verify our method only on CT images. In the future, we will verify our method on other modality images, such as MRI images, and we will extend the proposed algorithm to the data augmentation method that can boost 3D medical image segmentation accuracy.

\ack This work was supported by the National Natural Science Foundation of China (61402181, 61502174), the Natural Science Foundation of Guangdong Province (2015A030313215, 2017A030313358, 2017A030313355), the Science and Technology Planning Project of Guangdong Province (2016A040403046), the Guangzhou Science and Technology Planning Project (201704030051), and the Fundamental Research Funds for the Central Universities (2019MS073).

\bibliography{ecai}

\begin{thebibliography}{10}

\bibitem{bao2019depth}
Wenbo Bao, Wei-Sheng Lai, Chao Ma, Xiaoyun Zhang, Zhiyong Gao, and Ming-Hsuan
  Yang.
\newblock Depth-aware video frame interpolation.
\newblock In {\em Proceedings of the IEEE Conference on Computer Vision and
  Pattern Recognition}, pages 3703--3712, 2019.

\bibitem{chen2019multi}
Shuai Chen, Gerda Bortsova, Antonio Garcia-Uceda Juarez, Gijs van Tulder, and
  Marleen de~Bruijne.
\newblock \text{Multi-Task} \text{Attention-Based} \text{Semi-Supervised}
  \text{Learning} for \text{Medical} \text{Image} \text{Segmentation}.
\newblock {\em arXiv preprint arXiv:1907.12303}, 2019.

\bibitem{ciresan2012deep}
Dan Ciresan, Alessandro Giusti, Luca~M Gambardella, and J{\"u}rgen Schmidhuber.
\newblock Deep neural networks segment neuronal membranes in electron
  microscopy images.
\newblock In {\em Advances in neural information processing systems}, pages
  2843--2851, 2012.

\bibitem{dice1945measures}
Lee~R Dice.
\newblock Measures of the amount of ecologic association between species.
\newblock {\em Ecology}, 26(3):297--302, 1945.

\bibitem{dosovitskiy2015flownet}
Alexey Dosovitskiy, Philipp Fischer, Eddy Ilg, Philip Hausser, Caner Hazirbas,
  Vladimir Golkov, Patrick Van Der~Smagt, Daniel Cremers, and Thomas Brox.
\newblock Flownet: Learning optical flow with convolutional networks.
\newblock In {\em Proceedings of the IEEE international conference on computer
  vision}, pages 2758--2766, 2015.

\bibitem{goldman2008principles}
Lee~W Goldman.
\newblock Principles of \text{CT}: multislice \text{CT}.
\newblock {\em Journal of nuclear medicine technology}, 36(2):57--68, 2008.

\bibitem{van20073d}
Tobias Heimann, Bram Van~Ginneken, Martin~A Styner, Yulia Arzhaeva, Volker
  Aurich, Christian Bauer, Andreas Beck, Christoph Becker, Reinhard Beichel,
  Gy{\"o}rgy Bekes, et~al.
\newblock Comparison and evaluation of methods for liver segmentation from
  \text{CT} datasets.
\newblock {\em IEEE transactions on medical imaging}, 28(8):1251--1265, 2009.

\bibitem{hoogi2019self}
Assaf Hoogi, Brian Wilcox, Yachee Gupta, and Daniel~L Rubin.
\newblock \text{Self-Attention} \text{Capsule} \text{Networks} for \text{Image}
  \text{Classification}.
\newblock {\em arXiv preprint arXiv:1904.12483}, 2019.

\bibitem{Ilg2017FlowNet}
Eddy Ilg, Nikolaus Mayer, Tonmoy Saikia, Margret Keuper, Alexey Dosovitskiy,
  and Thomas Brox.
\newblock Flownet 2.0: Evolution of optical flow estimation with deep networks.
\newblock In {\em Proceedings of the IEEE conference on computer vision and
  pattern recognition}, pages 2462--2470, 2017.

\bibitem{jiang2018super}
Huaizu Jiang, Deqing Sun, Varun Jampani, Ming-Hsuan Yang, Erik Learned-Miller,
  and Jan Kautz.
\newblock Super slomo: High quality estimation of multiple intermediate frames
  for video interpolation.
\newblock In {\em Proceedings of the IEEE Conference on Computer Vision and
  Pattern Recognition}, pages 9000--9008, 2018.

\bibitem{jo2018quantitative}
YoungJu Jo, Hyungjoo Cho, Sang~Yun Lee, Gunho Choi, Geon Kim, Hyun-seok Min,
  and YongKeun Park.
\newblock Quantitative phase imaging and artificial intelligence: a review.
\newblock {\em IEEE Journal of Selected Topics in Quantum Electronics},
  25(1):1--14, 2018.

\bibitem{ali_emre_kavur_2019_3431873}
Ali~Emre Kavur, M.~Alper Selver, Oğuz Dicle, Mustafa Barış, and N.~Sinem
  Gezer.
\newblock {CHAOS - Combined (CT-MR) Healthy Abdominal Organ Segmentation
  Challenge Data}, 2019.

\bibitem{kingma2014adam}
Diederik~P Kingma and Jimmy Ba.
\newblock Adam: A method for stochastic optimization.
\newblock {\em arXiv preprint arXiv:1412.6980}, 2014.

\bibitem{kumar2013automatic}
SS~Kumar, RS~Moni, and J~Rajeesh.
\newblock Automatic liver and lesion segmentation: a primary step in diagnosis
  of liver diseases.
\newblock {\em Signal, Image and Video Processing}, 7(1):163--172, 2013.

\bibitem{lecun1995convolutional}
Yann LeCun, Yoshua Bengio, et~al.
\newblock Convolutional networks for images, speech, and time series.
\newblock {\em The handbook of brain theory and neural networks},
  3361(10):1995, 1995.

\bibitem{lecun1998gradient}
Yann LeCun, L{\'e}on Bottou, Yoshua Bengio, Patrick Haffner, et~al.
\newblock Gradient-based learning applied to document recognition.
\newblock {\em Proceedings of the IEEE}, 86(11):2278--2324, 1998.

\bibitem{liu2019end}
Shikun Liu, Edward Johns, and Andrew~J Davison.
\newblock End-to-end multi-task learning with attention.
\newblock In {\em Proceedings of the IEEE Conference on Computer Vision and
  Pattern Recognition}, pages 1871--1880, 2019.

\bibitem{liu2017video}
Ziwei Liu, Raymond~A Yeh, Xiaoou Tang, Yiming Liu, and Aseem Agarwala.
\newblock Video frame synthesis using deep voxel flow.
\newblock In {\em Proceedings of the IEEE International Conference on Computer
  Vision}, pages 4463--4471, 2017.

\bibitem{long2015fully}
Jonathan Long, Evan Shelhamer, and Trevor Darrell.
\newblock Fully convolutional networks for semantic segmentation.
\newblock In {\em Proceedings of the IEEE conference on computer vision and
  pattern recognition}, pages 3431--3440, 2015.

\bibitem{ma2018gan}
Shuang Ma, Jianlong Fu, Chang Wen~Chen, and Tao Mei.
\newblock \text{DA-GAN}: Instance-level image translation by deep attention
  generative adversarial networks.
\newblock In {\em Proceedings of the IEEE Conference on Computer Vision and
  Pattern Recognition}, pages 5657--5666, 2018.

\bibitem{masci2011stacked}
Jonathan Masci, Ueli Meier, Dan Cire{\c{s}}an, and J{\"u}rgen Schmidhuber.
\newblock Stacked convolutional auto-encoders for hierarchical feature
  extraction.
\newblock In {\em International Conference on Artificial Neural Networks},
  pages 52--59. Springer, 2011.

\bibitem{mavska2013segmentation}
Martin Ma{\v{s}}ka, Ond{\v{r}}ej Dan{\v{e}}k, Saray Garasa, Ana Rouzaut, Arrate
  Mu{\~n}oz-Barrutia, and Carlos \text{Ortiz-de-Solorzano}.
\newblock Segmentation and shape tracking of whole fluorescent cells based on
  the \text{Chan--Vese} model.
\newblock {\em IEEE transactions on medical imaging}, 32(6):995--1006, 2013.

\bibitem{mnih2014recurrent}
Volodymyr Mnih, Nicolas Heess, Alex Graves, et~al.
\newblock Recurrent models of visual attention.
\newblock In {\em Advances in neural information processing systems}, pages
  2204--2212, 2014.

\bibitem{niklaus2018context}
Simon Niklaus and Feng Liu.
\newblock Context-aware synthesis for video frame interpolation.
\newblock In {\em Proceedings of the IEEE Conference on Computer Vision and
  Pattern Recognition}, pages 1701--1710, 2018.

\bibitem{niklaus2017videof}
Simon Niklaus, Long Mai, and Feng Liu.
\newblock Video frame interpolation via adaptive convolution.
\newblock In {\em Proceedings of the IEEE Conference on Computer Vision and
  Pattern Recognition}, pages 670--679, 2017.

\bibitem{niklaus2017video}
Simon Niklaus, Long Mai, and Feng Liu.
\newblock Video frame interpolation via adaptive separable convolution.
\newblock In {\em Proceedings of the IEEE International Conference on Computer
  Vision}, pages 261--270, 2017.

\bibitem{oda2011organ}
Masahiro Oda, Teruhisa Nakaoka, Takayuki Kitasaka, Kazuhiro Furukawa, Kazunari
  Misawa, Michitaka Fujiwara, and Kensaku Mori.
\newblock Organ segmentation from \text{3D} abdominal \text{CT} images based on
  atlas selection and graph cut.
\newblock In {\em International MICCAI Workshop on Computational and Clinical
  Challenges in Abdominal Imaging}, pages 181--188. Springer, 2011.

\bibitem{oliveira2017augmenting}
Americo Oliveira, S{\'e}rgio Pereira, and Carlos~A Silva.
\newblock Augmenting data when training a \text{CNN} for retinal vessel
  segmentation: How to warp?
\newblock In {\em 2017 IEEE 5th Portuguese Meeting on Bioengineering (ENBENG)},
  pages 1--4. IEEE, 2017.

\bibitem{olshausen1993neurobiological}
Bruno~A Olshausen, Charles~H Anderson, and David~C Van~Essen.
\newblock A neurobiological model of visual attention and invariant pattern
  recognition based on dynamic routing of information.
\newblock {\em Journal of Neuroscience}, 13(11):4700--4719, 1993.

\bibitem{pereira2016brain}
S{\'e}rgio Pereira, Adriano Pinto, Victor Alves, and Carlos~A Silva.
\newblock Brain tumor segmentation using convolutional neural networks in
  \text{MRI} images.
\newblock {\em IEEE transactions on medical imaging}, 35(5):1240--1251, 2016.

\bibitem{ranjan2017optical}
Anurag Ranjan and Michael~J Black.
\newblock Optical flow estimation using a spatial pyramid network.
\newblock In {\em Proceedings of the IEEE Conference on Computer Vision and
  Pattern Recognition}, pages 4161--4170, 2017.

\bibitem{ronneberger2015u}
Olaf Ronneberger, Philipp Fischer, and Thomas Brox.
\newblock U-net: Convolutional networks for biomedical image segmentation.
\newblock In {\em International Conference on Medical image computing and
  computer-assisted intervention}, pages 234--241. Springer, 2015.

\bibitem{roth2015deeporgan}
Holger~R Roth, Le~Lu, Amal Farag, Hoo-Chang Shin, Jiamin Liu, Evrim~B Turkbey,
  and Ronald~M Summers.
\newblock Deeporgan: Multi-level deep convolutional networks for automated
  pancreas segmentation.
\newblock In {\em International conference on medical image computing and
  computer-assisted intervention}, pages 556--564. Springer, 2015.

\bibitem{sandfort2019data}
V~Sandfort, K~Yan, PJ~Pickhardt, and RM~Summers.
\newblock Data augmentation using generative adversarial networks
  (\text{CycleGAN}) to improve generalizability in \text{CT} segmentation
  tasks.
\newblock {\em Scientific reports}, 9(1):16884--16884, 2019.

\bibitem{simard2003best}
Patrice~Y Simard, David Steinkraus, John~C Platt, et~al.
\newblock Best practices for convolutional neural networks applied to visual
  document analysis.
\newblock In {\em ICDAR}, volume~3, 2003.

\bibitem{zhao2019data}
Amy Zhao, Guha Balakrishnan, Fredo Durand, John~V Guttag, and Adrian~V Dalca.
\newblock Data augmentation using learned transformations for one-shot medical
  image segmentation.
\newblock In {\em Proceedings of the IEEE Conference on Computer Vision and
  Pattern Recognition}, pages 8543--8553, 2019.

\bibitem{zhou2016view}
Tinghui Zhou, Shubham Tulsiani, Weilun Sun, Jitendra Malik, and Alexei~A Efros.
\newblock View synthesis by appearance flow.
\newblock In {\em European conference on computer vision}, pages 286--301.
  Springer, 2016.

\end{thebibliography}
\end{document}